\documentclass{aa}  

\usepackage{graphicx}
\usepackage[varg]{newtx}
\usepackage{amsmath}	
\usepackage{float}
\usepackage{geometry}
\usepackage{booktabs}
\usepackage{lscape}
\usepackage{threeparttable}
\usepackage{xcolor}
\usepackage{ulem}
\usepackage{cancel}
\usepackage{subcaption}

\newcommand{\nickel}{\ensuremath{^{56}\mathrm{Ni}}}

\newcommand{\artis}{\textsc{artis}}

\newcommand{\msun}{$M_\odot$}

\usepackage{natbib,twoopt}
\usepackage[breaklinks=true]{hyperref}
\bibpunct{(}{)}{;}{a}{}{,}             
\makeatletter
  \newcommandtwoopt{\citeads}[3][][]{\href{http://adsabs.harvard.edu/abs/#3}%
    {\def\hyper@linkstart##1##2{}%
     \let\hyper@linkend\@empty\citealp[#1][#2]{#3}}}
  \newcommandtwoopt{\citepads}[3][][]{\href{http://adsabs.harvard.edu/abs/#3}%
    {\def\hyper@linkstart##1##2{}%
     \let\hyper@linkend\@empty\citep[#1][#2]{#3}}}
  \newcommandtwoopt{\citetads}[3][][]{\href{http://adsabs.harvard.edu/abs/#3}%
    {\def\hyper@linkstart##1##2{}%
     \let\hyper@linkend\@empty\citet[#1][#2]{#3}}}
  \newcommandtwoopt{\citeyearads}[3][][]%
    {\href{http://adsabs.harvard.edu/abs/#3}
    {\def\hyper@linkstart##1##2{}%
     \let\hyper@linkend\@empty\citeyear[#1][#2]{#3}}}
\makeatother

\begin{document} 

   \title{NLTE spectral modelling for a carbon-oxygen and helium white-dwarf merger as a Ca-rich transient candidate}

   \titlerunning{NLTE spectral modelling for a CO+He WD merger for Ca-rich transients}

   \author{F. P. Callan
          \inst{1},
          A.~Holas\inst{2},
          J.~Mor\'an-Fraile\inst{2},
          S.~A.~Sim\inst{1,3,4},
          C.~E.~Collins\inst{5},
          L.~J.~Shingles\inst{6},
          J.~M.~Pollin\inst{1},
          F.~K.~R\"opke\inst{2,7,8},
          R.~Pakmor\inst{9}
          \and
          F.~R.~N.~Schneider\inst{2,7}
          }
   \authorrunning{F. P. Callan et al.}
   \institute{School of Mathematics and Physics, Queen's University Belfast, University   Road, Belfast BT7 1NN, UK \\
              \email{f.callan@qub.ac.uk} 
              \and
                       Heidelberger Institut f\"ur Theoretische Studien, Schloss-Wolfsbrunnenweg 35, D-69118, Heidelberg, Germany
               \and
Cosmic Dawn Center (DAWN), Denmark
               \and
Niels Bohr Institute, University of Copenhagen, Jagtvej 155A, DK-2200, Copenhagen N, Denmark
               \and
                       School of Physics, Trinity College Dublin, The University of Dublin, Dublin 2, Ireland
               \and
                       GSI Helmholtzzentrum f\"{u}r Schwerionenforschung, Planckstraße 1, 64291 Darmstadt, Germany
               \and
                       Zentrum f{\"u}r Astronomie der Universit{\"a}t Heidelberg, Astronomisches Rechen-Institut, M{\"o}nchhofstra{\ss}e 12--14, 69120 Heidelberg, Germany
               \and
                       Zentrum f{\"u}r Astronomie der Universit{\"a}t Heidelberg, Institut f{\"u}r Theoretische Astrophysik, Philosophenweg 12, 69120 Heidelberg, Germany
               \and
                       Max-Planck-Institut f\"{u}r Astrophysik, Karl-Schwarzschild-Str. 1, D-85748, Garching, Germany    
                }


\abstract{We carry out NLTE (non local thermodynamic equilibrium) radiative transfer simulations to determine whether explosion during the merger of a carbon-oxygen (CO) white dwarf (WD) with a helium (He) WD can reproduce the characteristic \ion{Ca}{ii}/[\ion{Ca}{ii}] and \ion{He}{i} lines observed in Ca-rich transients. Our study is based on a 1D representation of a hydrodynamic simulation of a $0.6\,$\msun\,CO\;+\;$0.4\,$\msun\,He WD merger. We calculate both photospheric and nebular-phase spectra including treatment for non-thermal electrons, as is required for accurate modelling of \ion{He}{i} and [\ion{Ca}{ii}]. Consistent with Ca-rich transients, our simulation predicts a nebular spectrum dominated by emission from [\ion{Ca}{ii}] 7291, 7324\,\AA\, and the \ion{Ca}{ii} near-infrared (NIR) triplet. The photospheric-phase synthetic spectrum also exhibits a strong \ion{Ca}{ii} NIR triplet, prominent optical absorption due to \ion{He}{i} 5876\,\AA\ and \ion{He}{i} 10830\,\AA\ in the NIR, as is commonly observed for Ca-rich transients. Overall, our results therefore suggest that CO+He WD mergers are a promising channel for Ca-rich transients. However, the current simulation overpredicts some \ion{He}{i} features, in particular both \ion{He}{i} 6678 and 7065\,\AA, and shows a significant contribution from \ion{Ti}{ii}, which results in a spectral energy distribution that is substantially redder than most Ca-rich transients at peak. Additionally the \ion{Ca}{ii} nebular emission features are too broad. Future work should investigate if these discrepancies can be resolved by considering full 3D models and exploring a range of CO+He WD binary configurations.

} 


   \maketitle

 \section{Introduction}
\label{sec:intro}
Ca-rich transients (see e.g.\,\citealt{taubenberger2017a} for a review) are a class of faint and fast-evolving supernovae (SNe) first observed in the early 2000s \citep{filippenko2003a}. They are primarily characterised by strong [\ion{Ca}{II}] emission and weaker [\ion{O}{I}] emission in their nebular spectra\footnote{A substantial amount of Ca is not necessarily required to excite the nebular \ion{Ca}{ii} lines \citep{shen2019a, jacobson-galan2020a, polin2019a}. However, we adopt ‘‘Ca-rich'' throughout as it is the naming that appears most commonly in the literature.} \citep{perets2010a, sullivan2011a, kasliwal2012a, valenti2014a, lunnan2017a, de2018a}. Other properties include peak luminosities that lie between novae and SNe ($\sim-14$ to $-17$ mag), fast photometric evolution and early evolution to the nebular phase \citep{kasliwal2012a}. Ca-rich transients show significant diversity in their spectroscopic properties with some more similar to Type Ia SNe and others more similar to Type Ibc SNe \citep{de2020a}.

The progenitors of Ca-rich transients are still unknown. However, the fact that they are typically found in remote locations in old host environments \citep{perets2010a, foley2015a, lunnan2017a} suggests older progenitors that involve white dwarfs (WDs) in binary systems. Various models have been proposed to explain Ca-rich transients including He shell detonations (e.g. \citealt{perets2010a, shen2010a, waldman2011a, woosley2011a}), tidal disruptions of white dwarfs either in a compact object binary (e.g. \citealt{margalit2016a, zenati2023a}) or by an intermediate mass black hole (e.g. \citealt{rosswog2009a, sell2015a}) and ultrastipped core-collapse supernovae (e.g. \citealt{kawabata2010a}) [see e.g. \citealt{shen2019a} for a review.]
 
In a previous work we presented a 3D hydrodynamic simulation of the merger of a 0.4\,\msun\ He WD and a 0.6\,\msun\ CO WD \citep{moran-fraile2024a}. 
During the merger, the He WD is tidally disrupted and the CO WD becomes engulfed in a massive  He ``shell''. As the orbit of the accreting material intersects with itself, efficiently transporting angular momentum outward, a region that is overdense relative to the surrounding gas develops. Compression and shear between this overdense region and the surface of the CO WD result in a He detonation ${\sim}$400\,s after the He WD was initially disrupted. The detonation wave propagates through the He shell that surrounds the CO WD, sending shocks inside it.
The convergence of these shocks ignites a C detonation that propagates outwards and burns the CO material.
Overall, the thermonuclear explosion resulting from this double detonation mechanism ejected 1\,\msun\ of material, synthesised 0.01\,\msun\ of \nickel\ and did not leave behind any remnant. Photospheric phase radiative transfer simulations of the model predicted a faint transient ($M_\mathrm{bol} \approx -15.7$) with light curves and spectra that resembled those of Ca-rich transients, including reproducing a prominent \ion{Ca}{II} near-infrared (NIR) triplet feature, as is observed in the photospheric spectra of such events. This suggests that such mergers are a possible progenitor system of Ca-rich transients. 

However, open questions remain. The \cite{moran-fraile2024a} radiative transfer simulation was unable to reproduce the strong feature around 5900\,\AA\, that is observed in the majority of Ca-rich transients, interpreted as either the \ion{He}{i} 5876\,\AA\ line or \ion{Na}{i} 5890, 5896\,\AA\ doublet (\ion{Na}{i} D). The lack of He features predicted, despite the significant mass of He in the model ejecta (0.3\,\msun), was not unexpected as the radiative transfer simulation utilised an approximate NLTE (non local thermodynamic equilibrium) treatment for the plasma conditions which did not include treatment for non-thermal electrons. As the temperatures of the ejecta of thermonuclear SN are insufficient to thermally excite \ion{He}{i}, it is these non-thermal electrons that dominate the population of the excited states of \ion{He}{i} such that spectral features can appear \citep{chugai1987a, lucy1991a, hachinger2012a}. The approximate NLTE treatment of the plasma conditions also means the simulation becomes increasingly unreliable beyond the photospheric phase. As such no predictions of the nebular phase spectra, by which Ca-rich transients are most readily identified, were possible from that simulation.

Here, we present full NLTE photospheric and nebular phase radiative transfer simulations, including treatment for non-thermal electrons, for a 1D spherical average of the $0.6\,$\msun\,CO+$0.4\,$\msun\,He WD merger model. In particular, our aim is to predict whether \ion{He}{i} lines form in the photospheric phase and if the nebular spectra display the prominent \ion{Ca}{ii} emission observed for Ca-rich transients.  
We describe our model ejecta structure and our radiative transfer set ups for the photospheric and nebular phase simulations in Section \ref{sec:num_methods}. We present our photospheric and nebular phase spectra in Section \ref{sec:results} before discussing the model predictions and comparing with observed Ca-rich transients in Section \ref{sec:comparisons_to_observations}. Finally, we present our conclusions in Section \ref{sec:conclusions}.

\section{Numerical methods}
\label{sec:num_methods}
\subsection{NLTE and non-thermal radiative transfer}
\label{subsec:RT_method}

The radiative transfer simulations are carried out using the time-dependent Monte Carlo radiative transfer code \artis\footnote{\href{https://github.com/artis-mcrt/artis/}{https://github.com/artis-mcrt/artis/}}~\citep{sim2007b, kromer2009a, bulla2015a, shingles2020a}. \textsc{artis} follows the methods of \citet{lucy2002a, lucy2003a, lucy2005a} and is based on dividing the radiation field into indivisible energy packet Monte Carlo quanta (packets hereafter). Here, we utilise the full NLTE and non-thermal capabilities added to \artis~by \citet{shingles2020a}. This includes a NLTE population and ionization solver and treatment for collisions with non-thermal leptons. These improvements both extend the validity of our radiative transfer simulations to the nebular phase and allow the spectral contribution of He to be predicted. To follow the energy distribution of high-energy leptons, which result from nuclear decays and Compton scattering of $\gamma$-rays, \artis~solves the Spencer-Fano equation (as framed by \citealt{kozma1992a}). Auger electrons are allowed to contribute to heating, ionization and excitation of bound electrons by non-thermal collisions is also included. \artis\ utilises the full Monte Carlo photon-packet trajectories to obtain a rate estimator for each photoionization cross-section included while a parameterized radiation field is used to estimate bound-bound transition rates. The atomic data sets we use are based on the compilation of \textsc{cmfgen} \citep{hillier1990a, hillier1998a} and are similar to that described by \cite{shingles2020a} but with some additional species and ions including atomic data for \ion{Ca}{IV}, \ion{Ti}{II}, \ion{Ti}{III}, \ion{Co}{i} and \ion{Ni}{i} from the most recent compilation of \textsc{cmfgen} (see \citealt{blondin2023a}).  In our photospheric phase simulation we include He \textsc{i-iii}, C \textsc{i-iv}, O \textsc{i-iv}, Na \textsc{i-iii}, Ne \textsc{i-iii}, Mg \textsc{i-iii}, Al \textsc{i-iv} Si \textsc{i-iv}, S \textsc{i-iv}, Ar \textsc{i-iv}, Ca \textsc{i-v}, Ti \textsc{ii-iv}, Cr \textsc{i-v}, Fe \textsc{i-v}, Co \textsc{ii-v} and Ni \textsc{ii-v} while we also include \ion{Co}{i} and \ion{Ni}{i} in our nebular phase simulation. We note that there is no \ion{Ti}{i} data in our atomic data set. We do not expect \ion{Ti}{i} to become the dominant ion in our simulations (we do include both \ion{Ca}{i} and \ion{Cr}{i}, which remain subdominant ions at all times). However, it is possible that \ion{Ti}{i} could contribute to the spectrum in the later phases we consider when the temperature and ionization state are low. However, we do not expect that this would significantly impact our principle conclusions on the formation of \ion{He}{i} and \ion{Ca}{ii} features. 

To cover both the photospheric and nebular phases we perform two separate simulations because a significantly greater number of packets are required when simulating the nebular phase (${\sim}$ factor of 40, see below). This is because we start with the packets representing gamma-rays and positrons arising from radioactive decays \citep{kromer2009a}. During the simulation these packets can be reprocessed to UVOIR wavelengths from which we ultimately extract the spectra. Owing to the optically thin conditions in the nebular phase, the fraction of the gamma-ray packets that are reprocessed is much smaller, meaning we require a much larger total number of packets in order to achieve good statistics in the required wavebands. Fortunately however, the number of interactions per packets in this optically thin regime is relatively small, meaning this increase in total number of packets does not present a computational problem.  

We utilised $5 \times 10^{7}$ packets for the photospheric phase simulation which describes the evolution from 1 to 40\,d post explosion with logarithmically spaced time steps ($\Delta \log (t) = 0.013$). The initial time steps were treated in local thermodynamic equilibrium (LTE) and at time step 12 (1.1\,d) the NLTE solver was switched on. We adopted a grey approximation in optically thick cells (those with a grey optical depth greater than 1000).

We initialised the nebular phase simulation at 25\,d post explosion and followed the spectroscopic evolution using $2 \times 10^{9}$ packets until 70\,d post explosion with logarithmically spaced time steps ($\Delta \log (t) = 0.0037$). The initial time steps of the nebular phase simulation were also treated in LTE and the NLTE solver was switched on at time step 12 (28\,d). 

\subsection{CO He WD merger ejecta model}
\label{subsec:ejecta_model}

\begin{figure}[htbp]
\centering
	\includegraphics[width=1.0\linewidth,trim={0cm 0cm 0cm 0cm},clip]{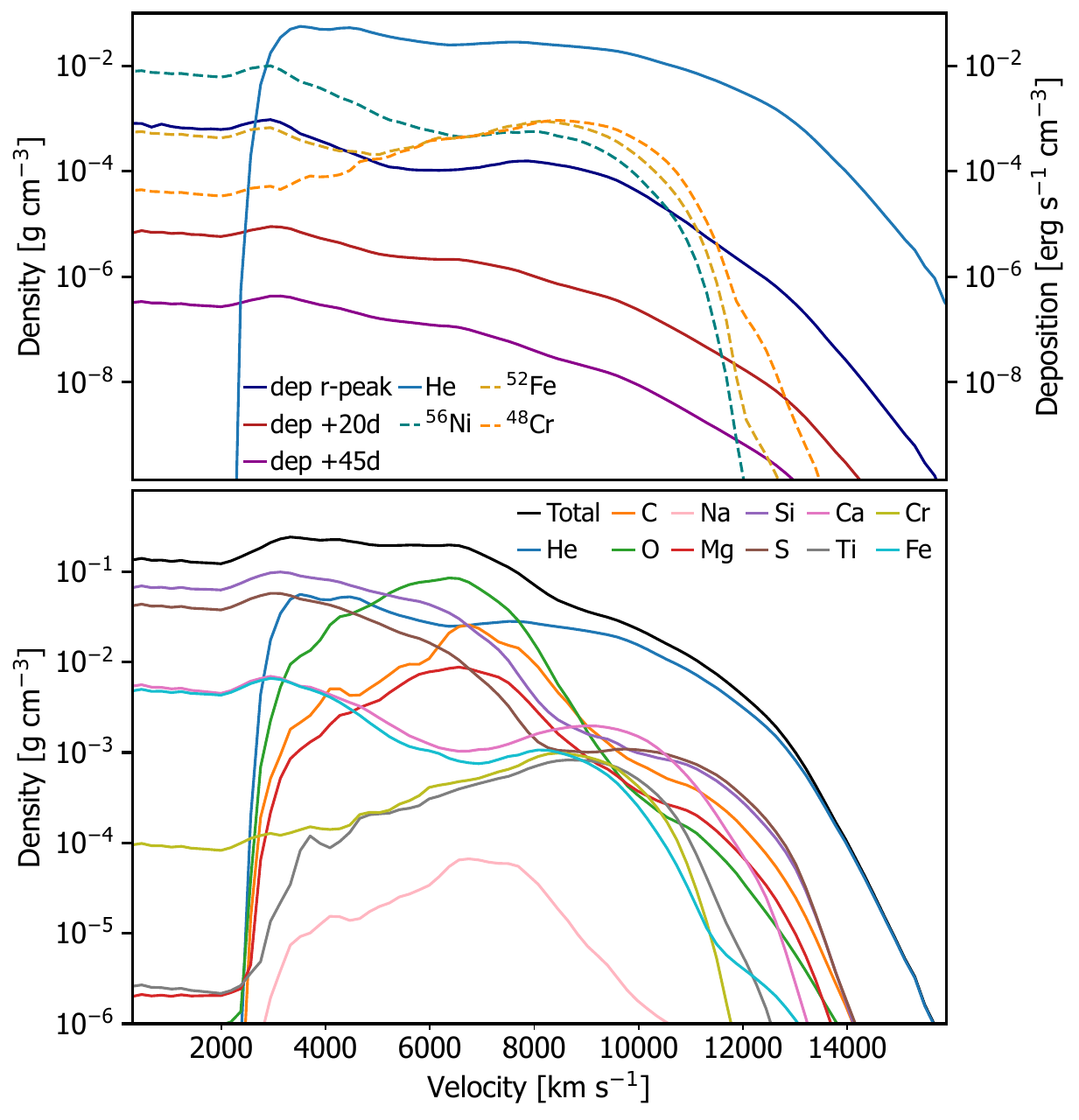}    
    \caption{\textit{Bottom panel:} model ejecta composition at 160\,s after explosion for key species in the simulation. \textit{Top panel:} energy deposition profiles for the three epochs we show spectra for. Overplotted for reference are the density profiles of He and key radioisotopes that we track the decay chain energy deposition for in the simulation.}
    \label{fig:Ejecta_composition}
\end{figure}

For this work, we utilise the WD merger model of \citet{moran-fraile2024a}, who simulated the merger of a 0.4\,\msun\ He WD and a 0.6\,\msun\ CO WD.
This simulation was carried out using the \textsc{arepo} code \citep{springel2010a,pakmor2011a,pakmor2016a,weinberger2020a}, which solves the ideal MHD equations on an unstructured moving Voronoi mesh using Newtonian self-gravity, including a Helmholtz equation of state \cite{timmes2000a}.
A 55 isotope nuclear reaction network, using the JINA reaction rates \cite{cyburt2010a} was coupled to the hydrodynamical solver.
The simulation was stopped when the ejecta reaches homologous expansion, around 100\,s after the explosion takes place.
Using a larger 384 isotope network, the $10^6$ Lagrangian tracer particles included in the simulation were post-processed to obtain precise nucleosynthetic yields \citep{seitenzahl2010a,pakmor2012a,seitenzahl2017a}.
For further details on the model, we refer the reader to \citet{moran-fraile2024a}.

\begin{table}[h!]
\centering
\caption{Compositions of key species in our radiative transfer input model in solar masses at 160\,s post explosion. Elemental masses are given in the upper part of table with the masses of radioisotopes whose decay chain energy deposition is tracked in the simulation included in the lower part of the table. 
}
\begin{tabular}{l c l c}
\hline
Element & Mass [$M_\odot$] \\
\hline
He   & $2.64 \times 10^{-1}$ \\
C    & $6.56 \times 10^{-2}$ \\
O    & $2.13 \times 10^{-1}$ \\
Ne   & $2.01 \times 10^{-2}$ \\
Na   & $2.08 \times 10^{-4}$ \\
Mg   & $2.76 \times 10^{-2}$ \\
Al   & $1.18 \times 10^{-3}$ \\
Si   & $1.70 \times 10^{-1}$ \\
S    & $8.18 \times 10^{-2}$ \\
Ar   & $1.50 \times 10^{-2}$ \\
Ca   & $1.91 \times 10^{-2}$ \\
Ti   & $5.16 \times 10^{-3}$ \\
Cr   & $5.43 \times 10^{-3}$ \\ 
Fe   & $1.03 \times 10^{-2}$ \\ 
Co   & $1.06 \times 10^{-3}$ \\ 
Ni   & $9.08 \times 10^{-3}$ \\ 
\hline
$^{48}$Cr  & $4.97 \times 10^{-3}$ \\
$^{52}$Fe  & $4.42 \times 10^{-3}$ \\
$^{56}$Ni  & $7.46 \times 10^{-3}$ \\
\hline
\label{tab:ejecta_composition}
\end{tabular}
\tablefoot{When mapping the output of the explosion simulation to our radiative transfer input model we exclude the outermost ejecta ($>16500 \,\mathrm{km}\,\mathrm{s}^{-1}$), which are very low density and thus will not contribute significant opacity. The yields listed here therefore show some differences to the yields that come directly from the nucleosynthesis post processing of the explosion simulation.}
\end{table}

While \artis~is capable of multi-dimensional simulations, owing to the increased computational cost of including our full NLTE treatment, we work in 1D for this first exploratory study and employ a spherically averaged model. The full 3D model has clear asymmetries (see \citealt{moran-fraile2024a} or \citealt{kozyreva2024a}) which likely impact the strengths and shapes of spectral features. However, the 1D average preserves the main characteristics of the model which allows us to address the key questions of whether \ion{He}{i} and [\ion{Ca}{ii}] line formation are plausible. We construct our 1D model by spherically averaging the 3D hydrodynamic simulation into 87 radial bins.
The composition of this 1D ejecta model is presented in Table\,\ref{tab:ejecta_composition} and shown in Figure~\ref{fig:Ejecta_composition} along with the energy deposition profiles at the epochs we present spectra for. From Figure~\ref{fig:Ejecta_composition} we can see that because the He distribution extends to low velocities there is significant overlap between where the He is present in the ejecta and where the deposition is taking place.  

\section{Results}
\label{sec:results}
Figures~\ref{fig:optical_emission_absorption} and \ref{fig:NIR_emission_absorption} show the optical and NIR spectra predicted by our simulations. We display three epochs with the r-max and +20\,d r-max spectra taken from our photospheric phase simulation and the +45\,d r-max spectrum taken from our nebular phase simulation\footnote{The spectra presented here will be made available on the Heidelberg supernova model archive HESMA (\citealt{kromer2017a}, {\href{https://hesma.h-its.org}{https://hesma.h-its.org}).}}.

\subsection{Optical spectroscopic evolution}
\label{subsec:optical_spectroscopic_evolution}

\begin{figure}[]
  \centering
  \includegraphics[width=.92\linewidth,trim={0.00 1.1cm 0 0},clip]{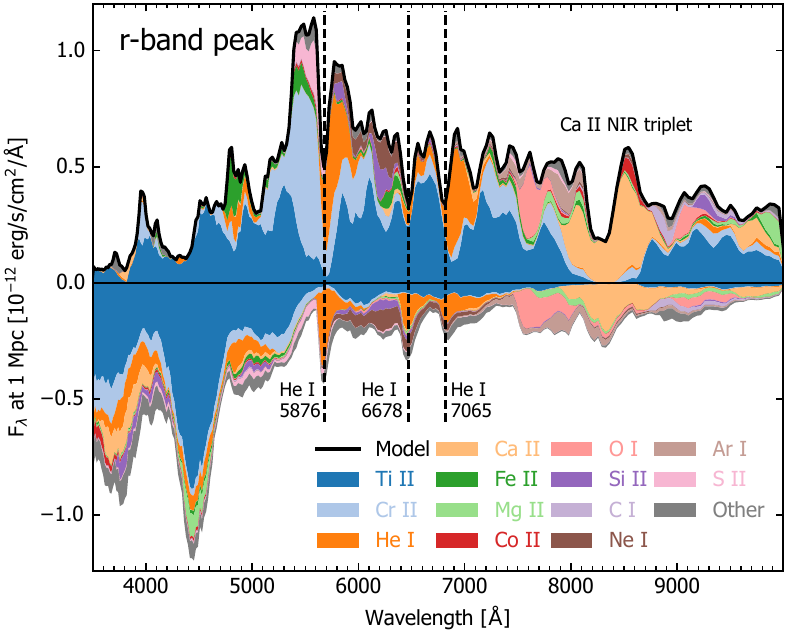}
  \includegraphics[width=.92\linewidth,trim={0.00 1.1cm 0 0},clip]{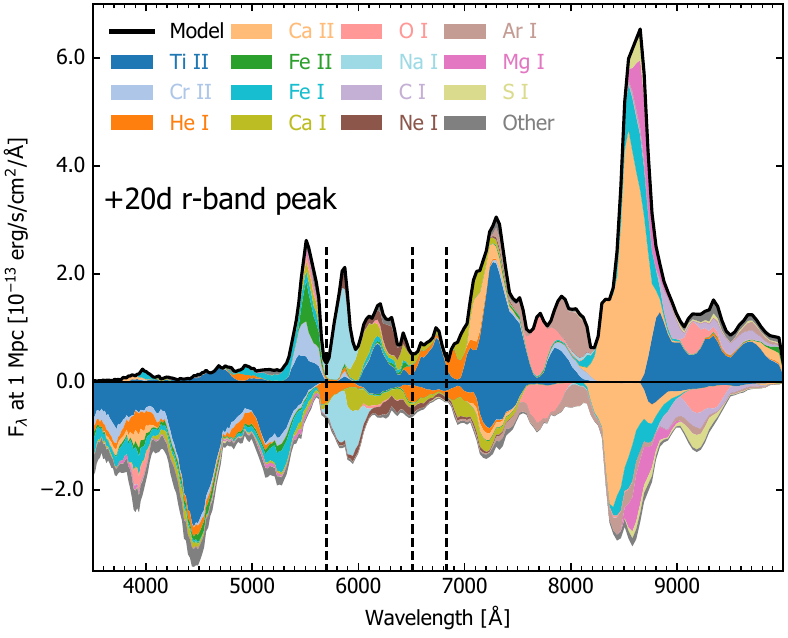}
  \includegraphics[width=.92\linewidth,trim={0.00 0.00cm 0 0},clip]{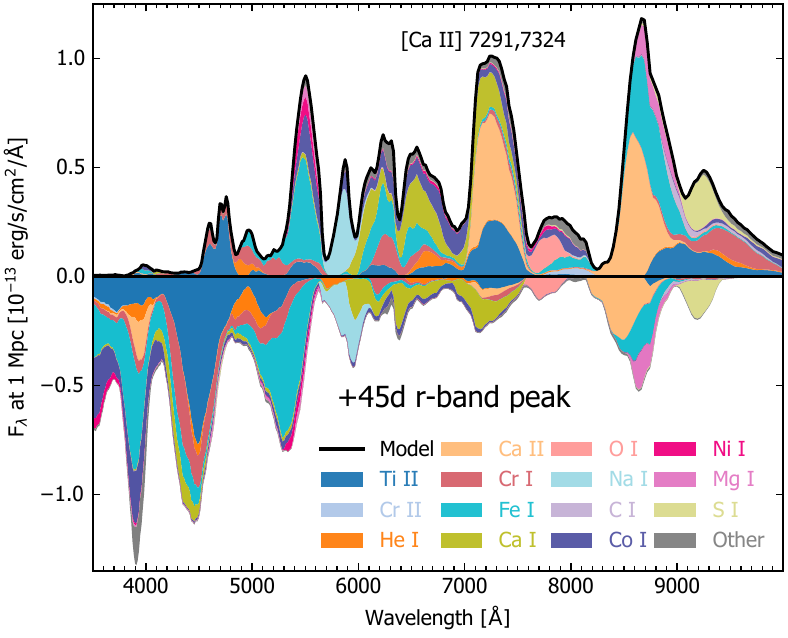}
  
  \caption{Optical spectroscopic evolution of our model from peak until the early nebular phase. The last species with which packets interacted before they left the simulation are indicated with different colours. Contributions to emission are plotted on the positive axis with contributions to absorption plotted on the negative axis. For reference, key lines are labelled along with their rest wavelength given in \AA. The dashed lines indicate the wavelengths of absorptions from the prominent optical \ion{He}{i} lines.
  }
  \label{fig:optical_emission_absorption}
\end{figure}  

The peak r-band optical spectrum of our simulation (see Figure\,\ref{fig:optical_emission_absorption}) predicts strong optical \ion{He}{i} features (\ion{He}{i} 5876, 6678 and 7065\,\AA) with \ion{He}{i} 5876\,\AA\ producing the strongest feature. This is a key difference to our previous radiative transfer simulation of the model \citep{moran-fraile2024a} which did not include treatment for non-thermal electrons and thus could not address whether \ion{He}{i} spectral features are expected to form. 

The \ion{Ca}{II} NIR triplet, which appears around 8300\,\AA, is the strongest individual feature in the peak spectrum. Additionally, there is a very significant contribution from \ion{Ti}{ii} and \ion{Cr}{ii} that covers almost the entire optical wavelength range. The substantial absorption, in particular from \ion{Ti}{ii}, blueward of ${\sim}$5000\,\AA\ and subsequent redistribution of this flux to redder wavelengths results in a relatively red spectral energy distribution (SED) at peak. There is a clear contribution from the \ion{O}{i} 7773\,\AA\ line at peak along with small contributions from intermediate mass elements (IMEs) such as Si and S. Only very weak contributions from iron group elements (IGEs) are predicted. 

At 20\,d post r-band peak, our simulation is already starting to show signs of a transition to the nebular phase with strong emission features typically associated with nebular phase spectra beginning to form. The spectrum still shows a very substantial contribution from \ion{Ti}{ii} while the spectral contribution of \ion{Cr}{ii} is significantly reduced. There is almost a complete suppression of flux blueward of ${\sim}$5000\,\AA\ due to line blanketing primarily from \ion{Ti}{ii}. The \ion{Ca}{ii} NIR triplet remains the strongest individual feature at this epoch showing very substantial emission. The simulation also predicts a strong emission feature around 7300\,\AA. Although this feature appears to be almost entirely dominated by \ion{Ti}{ii} 7355\,\AA\ with a small contribution to emission from forbidden [\ion{Ca}{ii}] 7291, 7324\,\AA\ (see Figure\,\ref{fig:optical_emission_absorption}), a significant amount of the underlying emission is actually attributed to [\ion{Ca}{ii}] with this emission then impacted by scattering and fluorescence from the \ion{Ti}{ii} 7355\,\AA\ line. 

The optical \ion{He}{i} features predicted in the peak spectrum are again present although the strength of the \ion{He}{i} lines are reduced relative to peak. In particular, the strong feature at ${\sim}$5700\,\AA\ which is almost entirely a result of \ion{He}{i} 5876\,\AA\ absorption in the peak spectrum is now a blend of \ion{He}{i} 5876\,\AA\ and \ion{Na}{i} D lines with \ion{Na}{i} contributing strongly to the formation of this feature, particularly in emission. This can primarily be attributed to variation in the ionization state of Na. At peak less than $10^{-6}$ of the total Na in the ejecta is \ion{Na}{i}. However, by 20\,d post peak this has increased to as much as $10^{-3}$ in certain parts of the model ejecta. Variation in the distribution of Na in our model ejecta (see Figure\,\ref{fig:Ejecta_composition}) may also be contributing, with Na only present in very small amounts in the very outer and inner ejecta but showing a significant increase in relative composition between ${\sim}$$3000 - 10000\,\mathrm{km}\,\mathrm{s}^{-1}$. As the photosphere recedes deeper into the ejecta with time, the spectrum forming region therefore evolves to overlap with this part of the ejecta that is richer in Na. There remains a relatively strong contribution from \ion{O}{i} 7773\,\AA\ at 20\,d post peak while there are still clear contributions from IMEs and only weak contributions from IGEs to the model spectrum.

By 45\,d post peak, our simulated optical spectrum becomes dominated by emission redward of 6000\,\AA\ although there is still significant absorption, particulary blueward of this, mainly due to \ion{Ti}{ii} and \ion{Fe}{i}. The most prominent spectral features are the \ion{Ca}{ii} NIR triplet and the forbidden [\ion{Ca}{ii}] 7291, 7324\,\AA\ emission feature which appear around 8700 and 7300\,\AA\ respectively. While the majority of the underlying emission for these features is from \ion{Ca}{ii} and [\ion{Ca}{ii}] respectively both features are still clearly impacted by scattering and fluorescence from allowed transitions (primarily \ion{Fe}{i}, \ion{Ti}{ii} and \ion{Mg}{i} for the NIR triplet and \ion{Ti}{ii} and \ion{Ca}{i} for the [\ion{Ca}{ii}] feature). 

\ion{Ti}{ii} remains a key species in the spectrum, contributing significantly across a range of wavelengths spanning the entire optical spectrum. There is a significant increase in the contributions of neutral species such as \ion{Fe}{i}, \ion{Ca}{i}, \ion{Cr}{i}, \ion{Si}{i} and \ion{Co}{i}. This is a result of the reduced ionization state of the simulation at this epoch. In particular, \ion{Fe}{i} has a substantial contribution across a wide range of optical wavelengths and the \ion{Fe}{i} feature predicted at ${\sim}5500$\,\AA, which also has contributions from \ion{Co}{i} and \ion{Ni}{i}, is the most prominent feature in the spectrum other than the \ion{Ca}{ii}/[\ion{Ca}{ii}] features discussed above. The complex of emission features predicted by the simulation between ${\sim}6000$-$7000$\,\AA\ is also primarily attributed to the blended contributions of neutral species (\ion{Fe}{i}, \ion{Ca}{i}, \ion{Cr}{i}, \ion{Co}{i}).

\subsection{NIR spectroscopic evolution}
\label{subsec:NIR_spectroscopic_evolution}

\begin{figure}
  \centering
  \includegraphics[width=.92\linewidth,trim={0.00 1.1cm 0 0},clip]{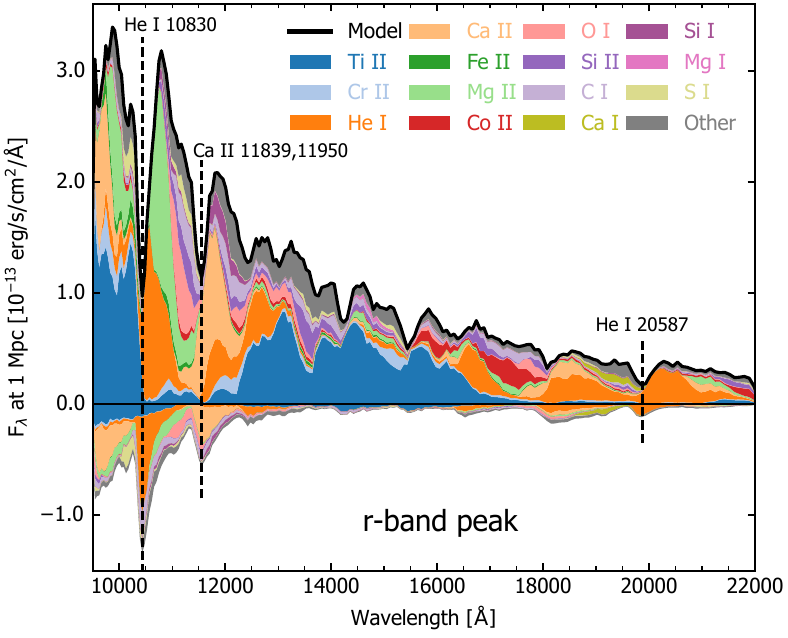}
  \includegraphics[width=.92\linewidth,trim={0.00 1.1cm 0 0},clip]{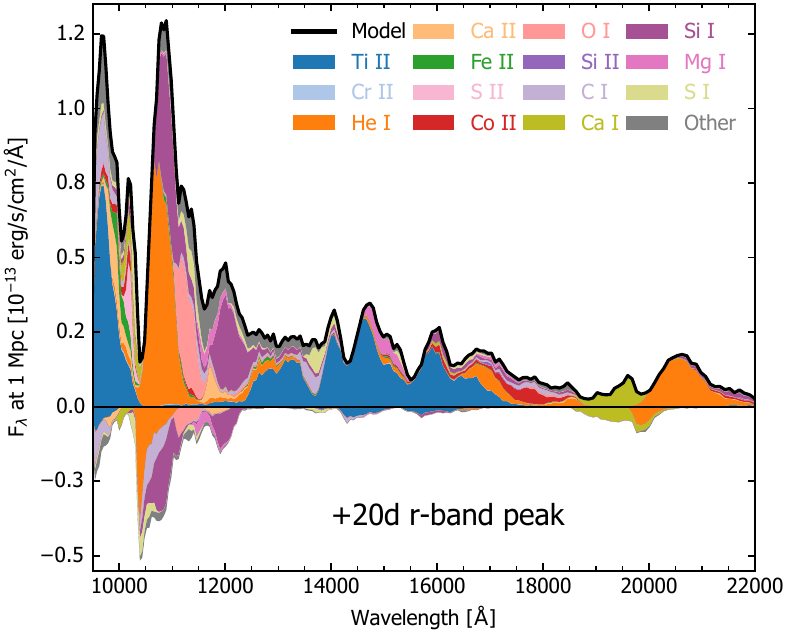}
  \includegraphics[width=.92\linewidth,trim={0.00 0.0cm 0 0},clip]{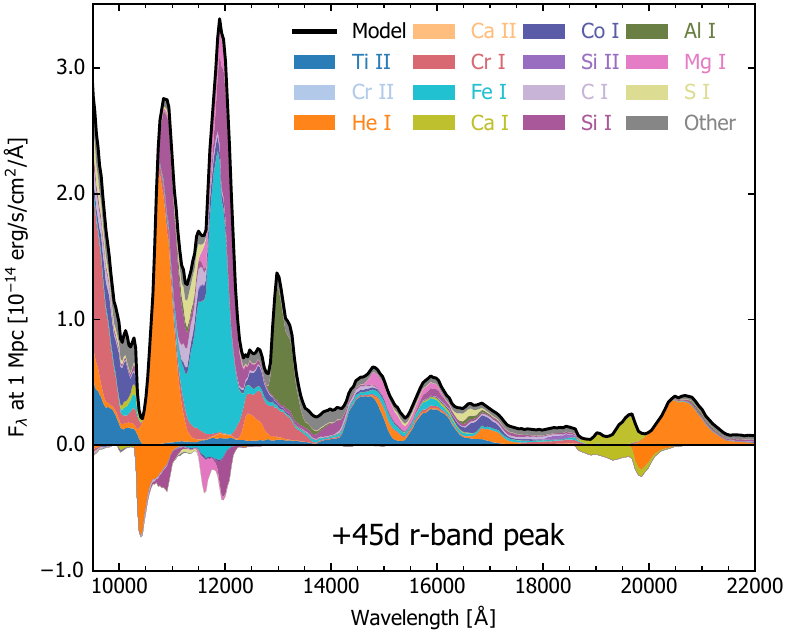}
  
  \caption{Same as Figure\,\ref{fig:optical_emission_absorption} for NIR spectroscopic evolution of the model.}
  \label{fig:NIR_emission_absorption}
\end{figure}  

The most prominent feature in the peak NIR spectrum predicted by our simulation is the strong \ion{He}{i} 10830\,\AA\ absorption feature (see Figure\,\ref{fig:NIR_emission_absorption}), while there is also a clear \ion{He}{i} 20587\,\AA\ absorption feature. We note that substantial ejected He masses, such as the ${\sim}0.3$\,\msun\ ejected for our model, are likely required for a \ion{He}{i} 20587\,\AA\ feature to appear: previous \artis\ simulations of CO WD explosions, ejecting ${\sim}0.04$\,\msun\ of He, predicted clear \ion{He}{i} 10830\,\AA\ features, but no obvious \ion{He}{i} 20587\,\AA\ feature formed \citep{collins2023a, callan2024b}. A prominent \ion{Ca}{ii} 11839, 11950\,\AA\ absorption feature is also present at ${\sim}11500$\,\AA. The NIR spectrum is dominated by \ion{Ti}{ii} emission for wavelengths shorter than ${\sim}$16000\,\AA\ while redward of this \ion{He}{i} provides the greatest contribution to the spectrum.

At 20 days post peak, the NIR spectrum starts to show emission features more typical of the nebular phase. \ion{He}{i} 10830\,\AA\ remains the strongest spectral feature predicted by the simulation in the NIR and the \ion{He}{i} 20587\,\AA\ feature is still clearly present. The \ion{Ca}{ii} absorption at ${\sim}11500$\,\AA\ has disappeared by this epoch but there is still significant emission from \ion{Ti}{ii}. There is also an increased contribution from neutral species, in particular \ion{Si}{i} and \ion{Ca}{i}.

At 45\,d post peak, clear \ion{He}{i} 10830 and 20587\,\AA\ features are still predicted despite no optical He features being present in the simulation by this epoch. The single strongest feature at this epoch is the strong emission feature at ${\sim}12000$\,\AA. This feature is primarily attributed to \ion{Fe}{i} but there is a also a contribution from \ion{Si}{i}. A prominent \ion{Al}{i} feature also appears at ${\sim}13500$\,\AA.

We note, for the small number of Ca-rich transients that have NIR spectral observations, a clear feature is present at wavelengths consistent with our predicted \ion{He}{i} 10830\,\AA\ feature \citep{valenti2014a, galbany2019a, jacobson-galan2020b, yadavalli2024a} for observations spanning from before peak until 40\,d post peak. Additionally, the Ca-rich transient SN~2019ehk \citep{jacobson-galan2020b} shows a feature consistent with the predicted \ion{He}{i} 20587\,\AA\ feature.  

\section{Comparison to observations}
\label{sec:comparisons_to_observations}
We now present spectral comparisons with observed Ca-rich transients focusing on the key \ion{Ca}{ii} and \ion{He}{i} features. Given that we are only comparing the observations to a single 1D model, and that there is  substantial spectral diversity displayed by Ca-rich transients (see e.g.\,\citealt{de2020a}), we cannot expect our current model to fully account for all features in the observations, or explain their variations. 
In the following we will however show that our simulations do predict the \ion{Ca}{ii}/[\ion{Ca}{ii}] and \ion{He}{i} features which are characteristic of Ca-rich transients.  

\subsection{Model comparisons with observations at peak}
\label{sec:peak_comparisons_to_observations}
\begin{figure}
  \centering
  \includegraphics[width=1\linewidth,trim={2.70cm 2.6cm 3cm 3cm},clip]{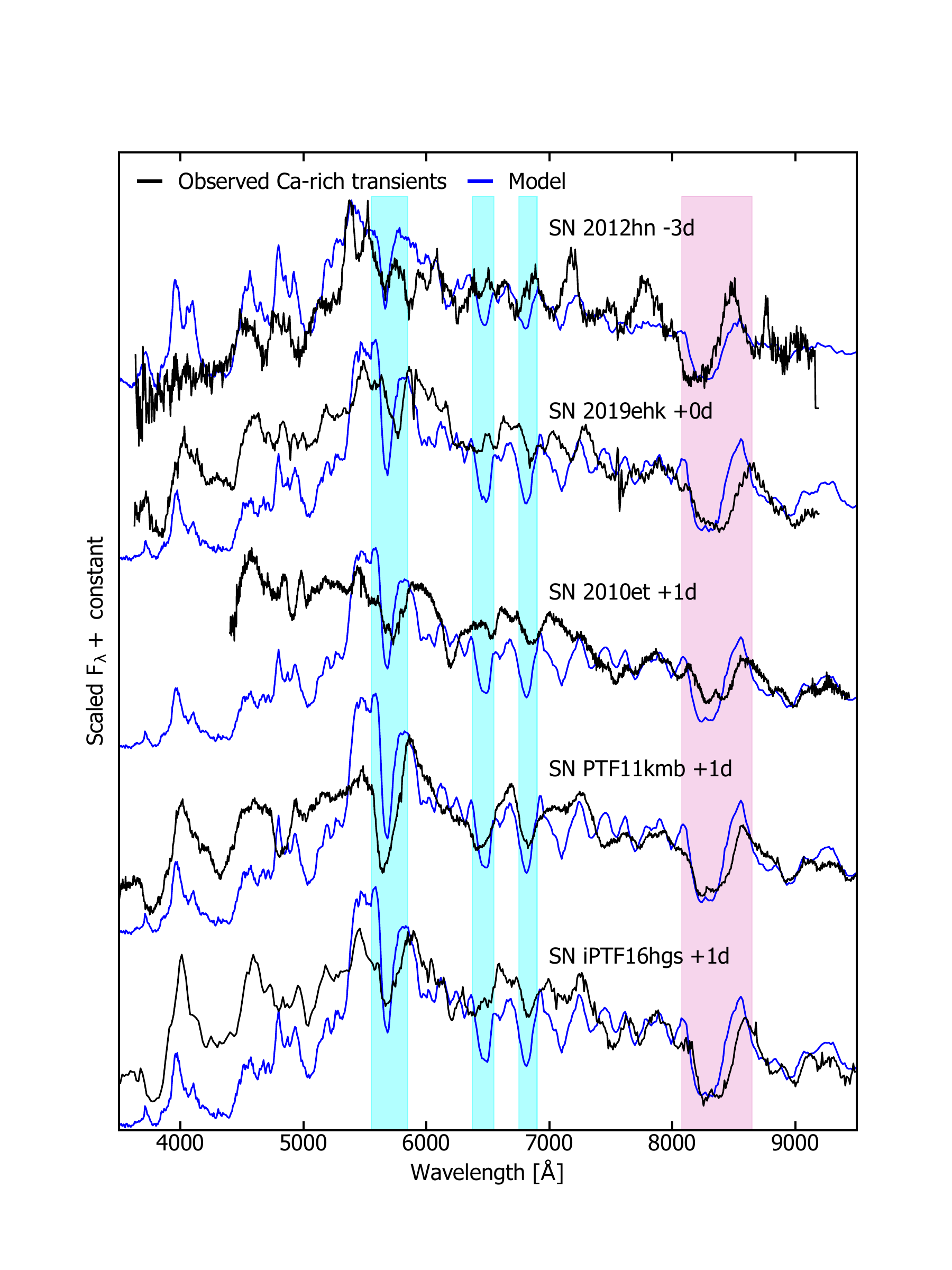}
  
  \caption{Spectroscopic comparisons around peak between our simulated spectra along with the observed Ca-rich transients  SN 2012hn \citep{valenti2014a}, SN 2019ehk \citep{jacobson-galan2020a,nakaoka2021a}, SN 2010et \citep{kasliwal2012a}, SN PTF11kmb \citep{lunnan2017a} and SN iPTF16hgs \citep{de2018a}. The model epochs are relative to the Sloan r-band peak and have been chosen to match the epochs of each observed spectrum relative to either Sloan r-band or Bessel R-band peak. For reference the strong optical \ion{He}{i} features, attributed to the \ion{He}{i} 5876, 6678 and 7065\,\AA\ lines, are highlighted in cyan and the \ion{Ca}{ii} NIR triplet is highlighted in pink.}
  \label{fig:peak_spectra_comparisons_with_observations}
\end{figure}
As can be seen from Figure\,\ref{fig:peak_spectra_comparisons_with_observations} there are very few spectral features at peak that are consistently present in the spectra of observed Ca-rich transients. There are, however, two features that are effectively ubiquitous in their peak optical spectra: the \ion{Ca}{ii} NIR triplet and the absorption feature attributed to either the \ion{He}{i} 5876\,\AA\ or \ion{Na}{i} D lines. Our simulation predicts a \ion{Ca}{ii} NIR triplet at peak with a strength and velocity generally consistent with that observed in the spectra of Ca-rich transients (although, as expected, there is variation in the quality of agreement depending on which Ca-rich transient we compare with). A clear \ion{He}{i} 5876\,\AA\ is also predicted, which has a strength comparable to the observed feature present around 5700\,\AA\ and also has a velocity which is in most cases consistent with that of the observed feature (but we note the deepest absorption of the predicted feature is at too high a velocity compared to the feature observed for SN~2019ehk). 

Our simulation predicts \ion{He}{i} 6678 and 7065\,\AA\ absorption features that are noticeably stronger than features at corresponding wavelengths in the observed Ca-rich transients we compare to in Figure\,\ref{fig:peak_spectra_comparisons_with_observations}, with the exception of SN PTF11kmb. The SED at peak predicted by our simulation is noticeably redder than the Ca-rich transients we compare to, other than SN~2012hn. As noted above, the red SED of our simulation is a result of significant absorption by \ion{Ti}{ii} at wavelengths bluer than ${\sim}5000$\,\AA\ and subsequent redistribution of this flux to redder wavelengths. The impact of \ion{Ti}{ii} in the spectra of Ca-rich transients shows substantial variation (see e.g.\,\citealt{de2020a}) with some events showing very little evidence of \ion{Ti}{ii} at peak while others show the same clear \ion{Ti}{ii} absorption trough at ${\sim}4300$\,\AA\ predicted by our simulation (see Figure\,\ref{fig:peak_spectra_comparisons_with_observations}). The Ti in our simulation is produced almost entirely in the He detonation. Therefore, although there are some Ca-rich transients with comparable \ion{He}{i} 6678 and 7065\,\AA\ line strengths and similarly red SEDs to our model, it is likely that the He (and Ti) mass of the model we investigate here is towards the high end of what would be expected for Ca-rich transients. We do however note that the He and Ti distributions vary strongly for different lines of sight as the ejecta show significantly asymmetric structures (see e.g. \citealt{kozyreva2024a} figures\,3 and 4). Such viewing angle effects are not captured by the spherically averaged model investigated here. Additionally, the strength of He and Ti features can be impacted by factors relevant to their ionization state such as the distribution of \nickel~in the ejecta which also shows significant variation with viewing angle in the 3D ejecta model. As such, future work is required to investigate whether different lines-of-sight in the 3D ejecta model can explain the variation in optical \ion{He}{i} and \ion{Ti}{ii} line strengths displayed by Ca-rich transients.   

\subsection{Model comparisons with observations in nebular phase}
\label{sec:nebular_comparisons_to_observations}
\begin{figure}
  \centering
  \includegraphics[width=1\linewidth,trim={2.70cm 2.7cm 2.0cm 4.4cm},clip]{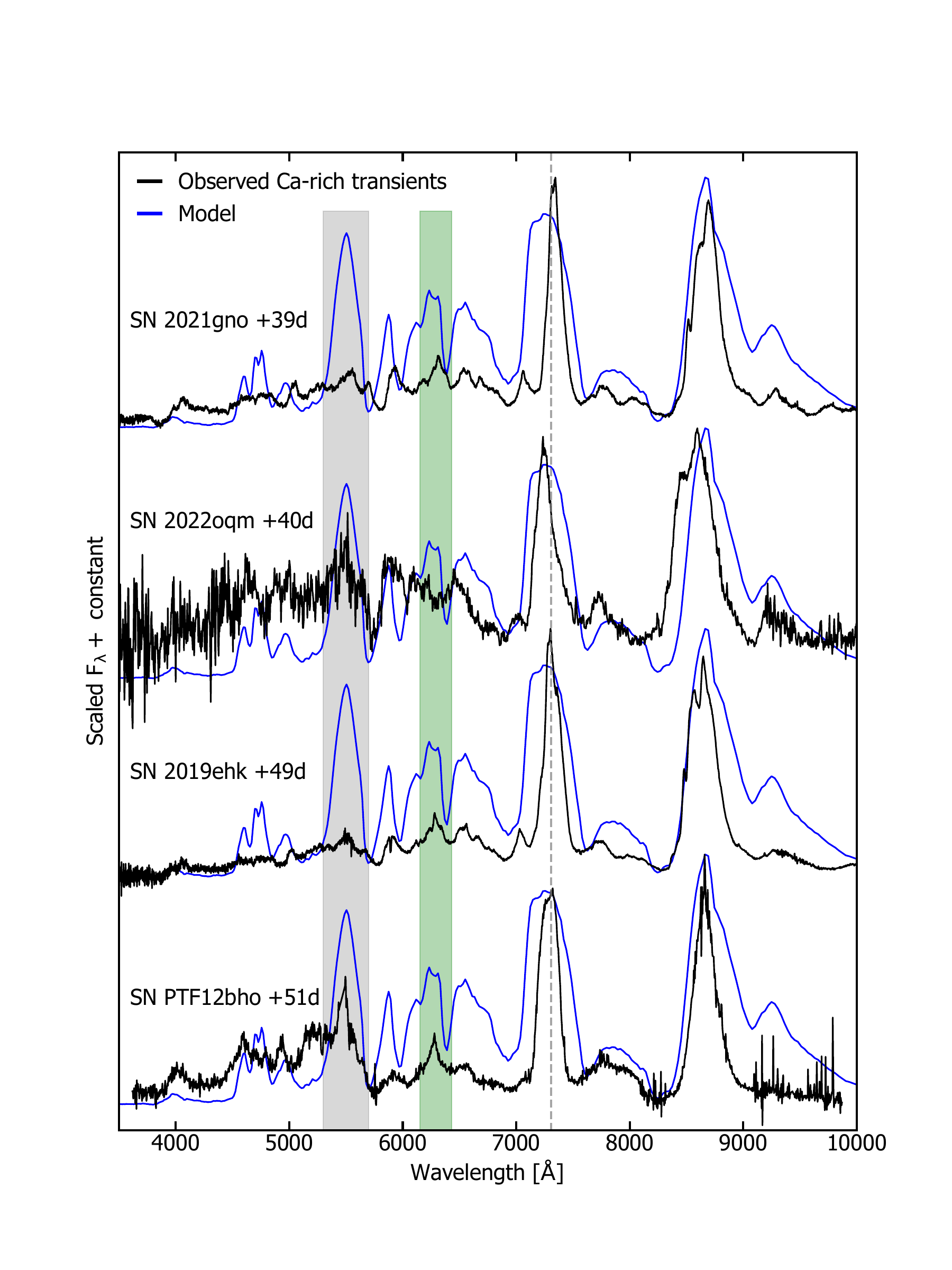}
  
  \caption{Nebular phase spectroscopic comparisons between our simulated spectra at 45\,d post r-band peak compared to the observed Ca-rich transients SN 2021gno \citep{jacobson-galan2022a, ertini2023a}, SN 2022oqm \citep{yadavalli2024a}, SN 2019ehk \citep{jacobson-galan2020a, nakaoka2021a} and SN PTF12bho \citep{lunnan2017a} at similar epochs. The epochs of the observations are relative to Sloan r-band or Bessel R-band peak. For reference, the grey dashed line is plotted between the rest wavelengths of the key [\ion{Ca}{ii}] 7291, 7324\,\AA\ lines, the \ion{Fe}{i} emission feature predicted by the simulation is highlighted in grey and the location of the [\ion{O}{i}] 6300, 6364\,\AA\ feature commonly observed for Ca-rich transients is highlighted in green.}
  \label{fig:nebular_spectra_comparisons_with_observations}
\end{figure}

As can be seen from Figure\,\ref{fig:nebular_spectra_comparisons_with_observations} the nebular phase spectrum predicted by our simulation is dominated by the strong [\ion{Ca}{ii}] and \ion{Ca}{ii} emission features at around 7300 and 8700\AA\  respectively. The simulated features have comparable strengths and velocities to the features observed in the nebular phase spectra of Ca-rich transients. However, our simulation predicts the \ion{Ca}{ii} NIR triplet emission to be stronger than the [\ion{Ca}{ii}] 7291, 7324\,\AA\ emission while these features have roughly equivalent strengths in the observed spectra\footnote{Radiative excitation still plays a role in the simulation at this time which keeps the NIR triplet relatively strong.}. Additionally, while we note the observed \ion{Ca}{ii} emission features show clear variations in their widths and velocities between different Ca-rich transients (see Figure\,\ref{fig:nebular_spectra_comparisons_with_observations}), our simulation predicts \ion{Ca}{ii} features that are too broad relative to the observations. 

From Figure\,\ref{fig:nebular_spectra_comparisons_with_observations} we can see that the [\ion{Ca}{ii}] emission feature shows at least some level of asymmetry for all the observed Ca-rich transients and in the case of SN 2022oqm even appears to show a blue shift. This suggests multi-dimensional structures are present in these observations. However, we note our 1D simulation also appears to show a blue shift in the [\ion{Ca}{ii}] feature but this is actually attributed to the feature being partially blended with \ion{Ti}{ii} and \ion{Ca}{i}. Future 3D radiative transfer simulations should therefore be carried out in order to investigate how the widths, velocity shifts and shapes of the \ion{Ca}{ii} features vary with viewing angle.  

Blueward of 7000\,\AA\ the observed Ca-rich transients show very substantial variation in their nebular phase spectra and as such the model has varying success in matching the observed spectra over this wavelength range. One feature that appears to be common to all the observations that we compare with here is the emission feature around 5500\,\AA, which the model attributes primarily to \ion{Fe}{i} emission. Other than SN 2022oqm, all the Ca-rich transients which we compare to predict an emission feature around 6300\,\AA\ attributed to  [\ion{O}{i}] 6300, 6364\,\AA. However, while the simulation predicts such an emission feature the contribution of [\ion{O}{i}] 6300, 6364\,\AA\ is quite minimal with blended emission from a variety of species including \ion{Fe}{i}, \ion{Cr}{i}, \ion{Ca}{i}, \ion{Co}{i} driving the formation of the feature (see Figure\,\ref{fig:optical_emission_absorption}). While this suggests the [\ion{O}{i}] 6300, 6364\,\AA\ feature is not a clean diagnostic at this epoch it will become more reliable at later times as forbidden emission increasingly dominates over the contribution of allowed transitions to the spectrum.

\section{Conclusions}
\label{sec:conclusions}
We have carried out 1D NLTE radiative transfer simulations, including treatment for non-thermal electrons, covering the photospheric and nebular phase for the merger of a 0.6\msun\ CO\;+\;0.4\msun\ He WD. Our aim was to establish whether the \ion{He}{i} and \ion{Ca}{ii}/[\ion{Ca}{ii}] features characteristic of Ca-rich transients are predicted, and we find that they are. In particular the simulation yields i) a nebular spectrum dominated by emission from [\ion{Ca}{ii}] 7291, 7324\,\AA\ and the \ion{Ca}{ii} NIR triplet, and ii) a peak spectrum which exhibits a strong \ion{Ca}{ii} NIR triplet and a prominent absorption feature at ${\sim}$5700\,\AA\ attributed to \ion{He}{i} 5876\,\AA\ in our simulation. 

Our simulation also predicts clear \ion{He}{i} 10830 and 20587\,\AA\ features which provide the best diagnostic of He in the model: in contrast to the optical He features, they are present at all epochs simulated, persisting into the nebular phase.
Although there are relatively few NIR observations of Ca-rich transients, existing observations do display a prominent feature at wavelengths similar to our predicted \ion{He}{i} 10830\,\AA\ feature \citep{valenti2014a, galbany2019a, jacobson-galan2020b, yadavalli2024a} for observations spanning from before peak until 40\,d post peak and SN~2019ehk also shows the \ion{He}{i} 20587\,\AA\ feature \citep{jacobson-galan2020b}. 
  
Therefore we conclude that low-mass CO WD$\;+\;$He WD mergers are promising candidates for Ca-rich transients. However, it is not yet clear whether such models can account for the full range of observed objects in this class of transients. Our model shows a strong spectral contribution from \ion{Ti}{ii} at all epochs resulting in an SED at peak that is substantially redder than the majority of Ca-rich transients. The strong [\ion{Ca}{ii}]/\ion{Ca}{ii} emission features that dominate our simulated nebular spectra are too broad relative to observations and the \ion{He}{i} 6678, 7065\,\AA\, absorption features predicted at peak appear to be more prominent than in the majority of Ca-rich transients. Interestingly the simulations presented by \cite{zenati2023a} for a CO WD disrupted by a hybrid HeCO WD also produce nebular spectra with strong [\ion{Ca}{ii}] but in contrast do not display any strong contributions from He. Therefore, taken together these different WD mergers can produce a family of transients that are unified by strong [\ion{Ca}{ii}] in their nebular spectra but show substantial variation in the strength of He spectral features predicted.

Future work should investigate whether the discrepancies between our model and observations can be resolved. In particular, the He and Ti distributions vary significantly with line-of-sight in the 3D hydrodynamic merger simulation, suggesting that observer orientation may be important. The strength of He and Ti features can also be impacted by factors relevant to their ionization state such as the distribution of \nickel~in the ejecta which also shows significant variation with viewing angle in the 3D ejecta model. Additionally, the [\ion{Ca}{ii}] feature can show asymmetric profiles and in some cases blue-shifts for observed Ca-rich transients, which may indicate that a multi-dimensional structure is present. Future multi-dimensional NLTE radiative transfer simulations should therefore be carried out to investigate if variation between lines-of-sight in the 3D ejecta model can better capture the variety observed for the Ca-rich class. 

Ca-rich transients show a variation of approximately three magnitudes in peak r-band brightness. Therefore, further hydrodynamic merger simulations should be carried out to explore if systems with different combinations of CO and He WD masses can robustly explode resulting in a diversity in peak brightness which can cover the variation observed for the Ca-rich class. Additionally, if systems with a lower mass He WD can explode, this should result in a reduction in the He and Ti ejected in the explosion. Future simultaneous optical and NIR observations (e.g.\,from SOXS \citealt{schipani2018a}) will be ideal for the testing of such models by capturing both the overall SED and the key \ion{He}{i} NIR features. 

\begin{acknowledgements}
FPC and SAS would like to acknowledge support from the UK Science and Technology Facilities Council (STFC, grant number ST/X00094X/1). This work used the DiRAC Memory Intensive service (Cosma8) at Durham University, managed by the Institute for Computational Cosmology on behalf of the STFC DiRAC HPC Facility (www.dirac.ac.uk). The DiRAC service at Durham was funded by BEIS, UKRI and STFC capital funding, Durham University and STFC operations grants. DiRAC is part of the UKRI Digital Research Infrastructure. The authors gratefully acknowledge the Gauss Centre for Supercomputing e.V. (www.gauss-centre.eu) for funding this project by providing computing time through the John von Neumann Institute for Computing (NIC) on the GCS Supercomputer JUWELS at Jülich Supercomputing Centre (JSC). AH is a fellow of the International Max Planck Research School for Astronomy and Cosmic Physics at the University of Heidelberg (IMPRS-HD) and acknowledges financial support from IMPRS-HD.     This work has received funding from the European Union’s Horizon Europe research and innovation programme under the Marie Skłodowska-Curie grant agreement No.~101152610. LJS acknowledges support by the European Research Council (ERC) under the European Union’s Horizon 2020 research and innovation program (ERC Advanced Grant KILONOVA No. 885281). JMP acknowledges the support of the Department for Economy (DfE). This work received support from the European Research Council (ERC) under the European Union’s Horizon 2020 research and innovation programme under grant agreement No.\ 759253 and 945806, the Klaus Tschira Foundation, and the High Performance and Cloud Computing Group at the Zentrum f{\"u}r Datenverarbeitung of the University of T{\"u}bingen, the state of Baden-W{\"u}rttemberg through bwHPC and the German Research Foundation (DFG) through grant no INST 37/935-1 FUGG. The authors acknowledge support by the state of Baden-Württemberg through bwHPC and the German Research Foundation (DFG) through grant INST 35/1597-1 FUGG. This work was supported by the Deutsche Forschungsgemeinschaft (DFG, German Research Foundation) -- RO 3676/7-1, project number 537700965, and by the European Union (ERC, ExCEED, project number 101096243). Views and opinions expressed are, however, those of the authors only and do not necessarily reflect those of the European Union or the European Research Council Executive Agency. Neither the European Union nor the granting authority can be held responsible for them. SAS thanks the Cosmic Dawn Center / Niels Bohr Institute (University of Copenhagen) who hosted his sabbatical visit during which part of this work was carried out. NumPy and SciPy \citep{oliphant2007a}, Matplotlib \citep{hunter2007a}  and \href{https://zenodo.org/records/15011510} {\textsc{artistools}}\footnote{\href{https://github.com/artis-mcrt/artistools/}{https://github.com/artis-mcrt/artistools/}} \citep{artistools_2025} were used for data processing and plotting.
\end{acknowledgements}

%
  \bibliographystyle{aa} 
  \bibliography{references} 
%

\end{document}